# GEANT4 Simulation of Nuclear Interaction Induced Soft Errors in Digital Nanoscale Electronics: Interrelation Between Proton and Heavy Ion Impacts

Artur M. Galimov[1,2], Regina M. Galimova[3], Gennady I. Zebrev[1*]

[1]*National Research Nuclear University MEPHI, Moscow, Russia*
[2]*MRI Progress, Moscow, Russia*
[3]*Kazan Federal University, Kazan, Russia*

**Abstract**

A simple and self-consistent approach has been proposed for simulation of the proton-induced soft error rate based on the heavy ion induced single event upset cross-section data and *vice versa*. The approach relies on the GEANT4 assisted Monte Carlo simulation of the secondary particle LET spectra produced by nuclear interactions. The method has been validated with the relevant in-flight soft error rate data for space protons and heavy ions. An approximate analytical relation is proposed and validated for a fast recalculation between the two types of experimental data.

## 1. Introduction

The susceptibility of digital integrated circuits (IC) to the proton-induced soft errors remains one of the most important problems of the microelectronics reliability. Due to low linear energy transfer (LET) the high energy space protons do not typically produce single event upsets (SEU) in the ICs via direct ionization. However, the proton-induced upsets can occur due to the ionization by the secondary products of the proton-induced nuclear interactions [1]. For example, the dominant abundance of energetic protons in space, especially in the South Atlantic Anomaly (SAA) region, makes them the main contributor to the soft error rate (SER) in spaceborne electronics [2]. A high level of proton flux is also a great problem for front-end electronics at the high-energy physics facilities [3, 4]. The situation with the proton-induced upsets gets worse in the modern commercial off-the-shelf (COTS) devices with nanoscale (< 100 nm) technology nodes. The multiple cell upsets dominate due to critical charge reduction in such circuits [5, 6]. The presence of the high-Z materials in the back-end-of-line (BEOL) materials also increases the IC vulnerability. The LETs of the secondary particles from the 500 MeV proton-W collisions may reach 34 MeV-cm$^2$/mg [7]. The particles with such high LET values can produce not only soft errors but also the destructive hard errors.

Modeling and prediction of the proton-induced SER in circuits requires generally the Monte Carlo simulation. Several Monte Carlo Geant4 based tools combined with the TCAD and SPICE (MRED [8], MUSCA SEP [9], TIARA [10], etc.) have been developed to simulate the processes at each level of radiation response of the ICs: the transport of radiation, the charge deposition and collection, the transistor and circuit response. These tools essentially require extremely detailed information about the internal structure of the IC. As practice shows, they are more convenient at the stage of the circuit design rather than for the SER prediction.

* Corresponding Author, tel. 7499-3240184
E-mail: gizebrev@mephi.ru.

In addition, as well as in the IRPP-based computational schemes [11], the traditional Monte Carlo based simulation toolkits are essentially based on the notions of the critical charge (energy) and the isolated sensitive volume (SV). We have noted in [12], that these concepts are violated at least for the highly scaled circuits due to non-local nature of the ionizing particle impact. We argue that an ion impact nonlocality transforms a set of the separate memory cell SVs into a single sensitive volume of the circuit as a whole. The concept of the node critical charge is a local circuit parameter which is more suitable for circuit simulations than for the ground test based SER prediction. We have proposed the approach which relies exceptionally on the phenomenological dependence of the SEU cross section on LET $\sigma(\Lambda)$ ($\Lambda$ is LET), which already contains all essential information about the SEU processes (generation, node collection, circuit response, etc.) [13]. This approach allows us to develop an effective Monte Carlo computational scheme for the nuclear interaction induced SEU prediction (see Sec. 2 of this paper).

Using a Geant4 toolkit, we have been simulated in this paper the LET spectra of the secondary particles (recoils and nuclear interaction products) in a single thin lamina IC sensitive volume followed by a SER calculation in the same way as for heavy ions. Notice that the nanoscale commercial circuits have a very small value of noise immunity characterized by a critical charge as low as parts of 1 fC [14]. A very low ionization and collection path of the secondary ion is needed to cause the cell upset(s). If the ranges of secondary particles are greater than a typical collection lengths, the sensitive volume could be considered as being exposed to an effective flux of the secondary particles with a given LET spectrum [15, 11, 16]. The LET spectrum of secondary particles in this case becomes a good metrics in describing the proton-induced SEUs [17, 18]. The problem of a use of the proton ground test data as a proxy for the heavy ion susceptibility also has a great practical value [19, 20]. We have proposed here a simplified approach to recover the ion-induced cross sections and SER in the commercial circuits based only on proton ground test (see Sec. 3).

## 2. General approach
### 2.1 Model outline

There are two types of the single event errors in microelectronic circuits, namely, the errors due to direct ionization of primary particles (mainly, the heavy ions with $Z \geq 2$), and errors due to ionization by the secondary products of elastic or non-elastic nuclear interactions (recoil atoms, fission or fusion) formed upon irradiation with the primary particles (mainly, by protons). Both types are determined ultimately by direct ionization with heavy ions of either primary or secondary origin. Therefore, we will consider here the LET spectra of the primary particles $\phi_{PRIM}(\Lambda)$ and the LET spectra of secondary products of nuclear interactions $\phi_{SEC}(\Lambda)$ on an equal footing. The former is usually given by the external conditions (e.g., the LET spectra of the heavy ions in space environments). The latter, on the contrary, depends not only on the external particle fluxes, but also on the cross sections of the nuclear interactions, and also on the integrated circuit characteristics (layout, chemical composition, overlayer structure etc.).

We assume that both types of SEUs are determined by the same LET dependent cross section $\sigma(\Lambda)$, which is a phenomenological circuit parameter characterizing the tolerance of the electronic component to an impact of the ionizing particle at a given LET. Then, the total SER can be calculated as follows

$$SER_{total} = \int \sigma(\Lambda)\left[\phi_{PRIM}(\Lambda) + \phi_{SEC}(\Lambda)\right]d\Lambda, \quad (1)$$

where $\sigma(\Lambda)$ is the ion-induced SEU cross-section averaged over all directions, $\phi_{PRIM}(\Lambda)$ is the omnidirectional differential LET spectrum of primary particles, $\phi_{SEC}(\Lambda)$ is the differential LET spectrum of secondary particles including the recoils and products of inelastic nuclear interactions. The first term in (1) is calculated straightforwardly, using the LET spectra, provided by a standard tool (see, e.g., [21]). GEANT4 can be used as an appropriate tool to simulate the secondary particle LET spectrum $\phi_{SEC}(\Lambda)$ at given proton energy spectra [22, 23, 24].

### 2.2 Details of GEANT4 computational procedures

Fig. 1 shows a multilayered planar structure of the IC overlayer which have been used in all simulation sets in this work. GEANT4 tools were used to calculate the LET spectra of secondary particles $\phi_{SEC}(\Lambda)$ in a thin (50 nm) virtual detector in the range $\Lambda$ greater than 0.5 MeV-cm$^2$/mg.

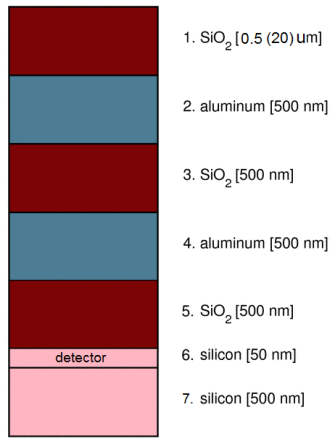

Fig. 1. Generic view of a simplified structure of the IC overlayer used at simulation.
Top oxide thicknesses at simulations were taken either 0.5 μm or 20 μm.

We consistently use the concept of thin lamina as a sensitive volume (SV) of the whole circuit. The width and length of virtual detector are chosen equal to 1 cm (conditionally speaking, the IC sizes). Unlike the conventional IRPP technique [20, 25], a use of such approach allows us reducing the fluence of the projectile protons to a level $10^8$ cm$^{-2}$ without an applying of biasing techniques [25].

Based on GEANT4 v.10.3 library [26, 27], a custom simulator of proton-induced interaction in the circuits has been developed. The structure of GEANT4 allows configuring easily the test target, surrounding radiation environment and processing physics lists. The physics list consists of native Geant4 classes, which partially are used in MRED [28]. They are the G4EmStandardPhysics (option 4) for electromagnetic processes and QGSP_BIC_HP physics list for the hadronic processes. Several works confirm an accuracy of the chosen physics lists [25, 29]. The GEANT4 General Particle Source (GPS) was used to model the radiation environment in space and under the accelerator. The OMERE tool [21] was used to generate the primary proton fluxes on a given orbit. Benchmarking simulations of $\phi_{SEC}(\Lambda)$ showed a good agreement with the similar results of Hiemstra et al. [30].

*2.3 General notions*

An incident flux of primary protons with energy $\varepsilon_p$ and differential energy spectrum $\phi_p(\varepsilon_p)$ produces the secondary ion flux with the LET spectrum $\phi(\Lambda,\varepsilon_p)$ [cm$^{-2}$s$^{-1}$/MeV-cm$^2$/mg/MeV]. We define the joint distribution $\phi(\Lambda,\varepsilon_p)$ through the two-variable conversion function $p(\Lambda|\varepsilon_p)$ as follows

$$\phi(\Lambda,\varepsilon_p) = p(\Lambda|\varepsilon_p)\phi_p(\varepsilon_p). \tag{2}$$

In fact, this relation is a Bayesian formula expressing the joint flux distribution $\phi(\Lambda,\varepsilon_p)$ via a conditional probability $p(\Lambda|\varepsilon_p)$ at a given proton energy spectrum $\phi_p(\varepsilon_p)$. Generally, the conditional probability distribution $p(\Lambda|\varepsilon_p)$ contains all information about transformations of the primary fluxes into the secondary ones, and it can be straightforwardly calculated with the Monte Carlo computational tools. This a central point of our approach. Notice that $p(\Lambda|\varepsilon_p)$ can be simulated either for a unidirectional proton beam, or for the omindirectional proton fluxes in space environments.

Full information obtaining requires the computationally intensive efforts. Therefore, a reduced description turns out to be useful in practice. For example, it is useful to define a differential secondary particle generation efficacy $\alpha(\varepsilon_p)$

$$\alpha(\varepsilon_p) = \int p(\Lambda|\varepsilon_p) d\Lambda = \frac{\int \phi(\Lambda,\varepsilon_p) d\Lambda}{\phi_p(\varepsilon_p)}. \tag{3}$$

The integral of the joint distribution $\phi(\Lambda,\varepsilon_p)$ over the both variables yields the full flux of secondary particles

$$\phi_{SEC} = \int d\varepsilon_p \int d\Lambda\, \phi(\Lambda,\varepsilon_p) = \int d\varepsilon_p\, \alpha(\varepsilon_p)\phi_p(\varepsilon_p) \equiv \alpha \int d\varepsilon_p\, \phi_p(\varepsilon_p), \tag{4}$$

where the integral generation efficacy $\alpha$ at a given proton energy spectrum is implicitly defined. The differential efficacy $\alpha(\varepsilon_p)$ was estimated in [11, 16] as follows

$$\alpha(\varepsilon_p) = \Sigma(\varepsilon_p) N_{at} L_R(\varepsilon_p), \tag{5}$$

where $N_{at}$ is the atom density in material (~5×10$^{22}$ cm$^{-3}$ in silicon), $\Sigma(\varepsilon_p)$ is the total cross section of nuclear interactions which is approximately equal in silicon to 1 - 2 barns for proton in the energy range from ten to hundreds MeV (see for example, [31])), $L_R(\varepsilon_p)$ is a mean range of secondary particles. Strictly speaking, the secondary particle generation efficacy $\alpha(\varepsilon_p)$ should also include the average number of the products per a nuclear reaction, but this uncertainty is absorbed by the uncertainty of the secondary particle ranges. Assuming $\Sigma(\varepsilon_p) \sim 2 \times 10^{-24}$ cm$^2$ and $L_R \sim$ 1 μm, one gets approximately $\alpha \cong 10^{-5}$. This is an order of magnitude of the ratio of the proton to the heavy ion SEU cross sections (typically, of order $10^{-13}$ and $10^{-8}$ cm$^2$, respectively).

This simple qualitative picture can be illustrated by the Monte Carlo calculation in which the generation efficacy, defined in (3), is simulated as a function of the primary proton energy for two different overlayer thicknesses. The simulation results are shown in Fig. 2.

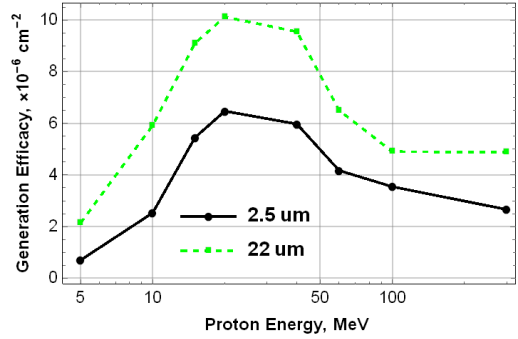

Fig. 2  The efficacy of secondary particle generation $\alpha(\varepsilon_p)$ simulated as a function of proton energies $\varepsilon_p$ for overlayer thickness 2.5 μm (solid lne) and 22 μm (dashed line).

Simulation results are in a good agreement with the estimation by (5) ($\alpha \sim 10^{-5}$). Larger overlayer thickness provides a larger generation efficacy in full accordance with (5). The remarkable non-monotonic form of the $\alpha(\varepsilon_p)$ dependence also has a simple qualitative explanation. The decrease in efficacy at large proton energies can be explained by the energy dependence of the elastic Rutherford scattering $\Sigma_{Rutherford}(\varepsilon_p) \propto 1/\varepsilon_p$, which is the dominant mechanism of nuclear interactions at relatively low energies. The efficacy at the low proton energies grows due to an increase in the range of secondary particles $L_R(\varepsilon_p)$.

### *2.4 Soft error rates from secondary particles*

The LET spectrum of the secondary particles can be expressed via (2) as follows

$$\phi_{SEC}(\Lambda) = \int \phi(\Lambda, \varepsilon_p) d\varepsilon_p = \int p(\Lambda | \varepsilon_p) \phi_p(\varepsilon_p) d\varepsilon_p . \tag{6}$$

Then, the soft error rate produced by secondary particles could be generally written as follows

$$SER = \int \sigma(\Lambda) \phi_{SEC}(\Lambda) d\Lambda = \int d\varepsilon_p \int d\Lambda \, \sigma(\Lambda) p(\Lambda | \varepsilon_p) \phi_p(\varepsilon_p) \tag{7}$$

This general relation can be represented in the two mathematically equivalent forms. Integrating first over $\varepsilon_p$, we have 'the LET representation' (1), while integrating over the entire LET range yields 'the energy representation' for the soft error rate

$$SER = \int \sigma_p(\varepsilon_p) \phi(\varepsilon_p) d\varepsilon_p , \tag{8}$$

where

$$\sigma_p(\varepsilon_p) = \int \sigma(\Lambda) p(\Lambda | \varepsilon_p) d\Lambda . \tag{9}$$

Defined in such way $\sigma_p(\varepsilon_p)$ should be identified with the phenomenological proton SEU cross section measured under the tests with a given proton energy. Thus, the notion of the conditional conversion function $p(\Lambda|\varepsilon_p)$ is able to represent a description of the SER from the primary and secondary particles in a unified way. The next sections are devoted to the two aspects of the problem.

### 3. Monte Carlo simulation and validation of the results

*3.1 Direct problem: proton induced SER from heavy ion tests*

Suppose we know the heavy ion induced SEU cross section $\sigma(\Lambda)$ and our task is to calculate the proton-induced SER in space or in other hazardous environment for a given proton energy spectrum $\phi(\varepsilon_p)$. Then we should to calculate with the GEANT4 the LET spectrum of secondary particles (6) followed by a use of (1). This method is also discussed by Tang in [1]. For a wide class of the commercial (COTS) nanoscale (feature nodes < 100 nm) circuits the above-threshold ($\Lambda > \Lambda_C$) ion cross section vs LET dependence can be approximated by a linear function [6, 12, 13]

$$\sigma(\Lambda) \cong K_d (\Lambda - \Lambda_C), \tag{10}$$

where $\Lambda_C$ is the critical (threshold) LET, and $K_d$ is the empirical slope of the SEU cross section dependence. We obtain in this case a simple relation for proton-induced SER

$$SER \cong K_d \phi_{SEC} [\langle\Lambda\rangle - \Lambda_C] = K_d \alpha \phi_p [\langle\Lambda\rangle - \Lambda_C], \tag{11}$$

where $\phi_{SEC}$ is the full flux of secondary particles $\phi_{SEC} = \int \phi_{SEC}(\Lambda) d\Lambda$. The LET, averaged over the whole secondary particle spectrum, referred here to as the LET spectrum rigidity

$$\langle\Lambda\rangle \equiv \frac{\int \Lambda \phi_{SEC}(\Lambda) d\Lambda}{\int \phi_{SEC}(\Lambda) d\Lambda}. \tag{12}$$

For monoenergetic proton flux typically used during ground tests we have $\phi_p(\varepsilon_p) \cong \phi_p \delta(\varepsilon'_p - \varepsilon_p)$ and the LET spectrum rigidity for a given energy of primary protons takes a form

$$\bar{\Lambda}(\varepsilon_p) = \frac{\int \Lambda p(\Lambda|\varepsilon_p) d\Lambda}{\alpha(\varepsilon_p)} \tag{13}$$

Figure 4 shows the rigidity of the secondary LET spectrum $\bar{\Lambda}(\varepsilon_p)$ simulated at normal incidence of the mono-directional proton beams in a wide range of energies.

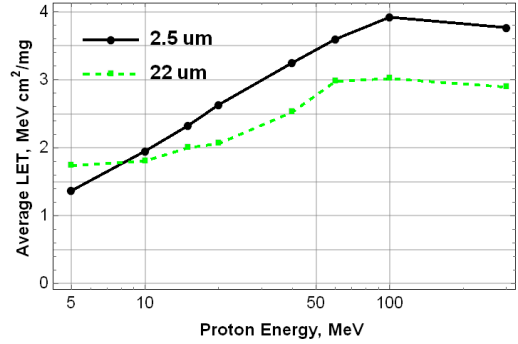

Fig. 3 The rigidity (average LET) of the secondary particle LET spectrum, simulated as a function of incident proton energy.

A non-monotonic behavior of the LET spectrum rigidity can be explained by the fact that the average energy of the secondary particles increases with proton energy, and the LETs of these particles decrease, following the well-known Bethe-Bloch expression.

Using (9) with the custom GEANT4 tools we simulated the cross section of proton-induced SEU $\sigma_p(\varepsilon_p)$ based on the typical values of heavy ion cross section parameters $K_d = 10^{-8}$ mg/MeV and $\Lambda_C = 0.5$ MeV-cm$^2$/mg. Figures 4 show the SEU cross sections for unidirectional protons simulated for different incident tilt angles.

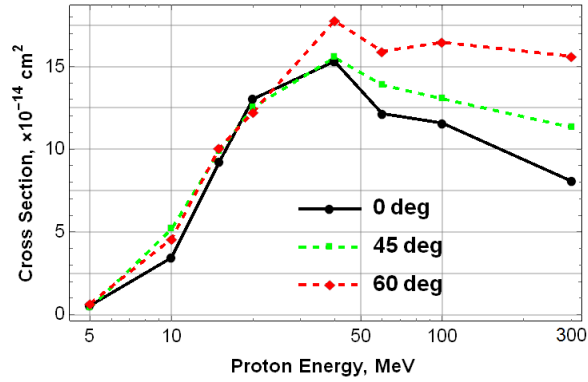

Fig. 4 The simulated curves $\sigma_P(\varepsilon_p)$ at three different tilt angles (0, 45 and 60 degrees). Heavy ion cross section parameters were supposed to be $K_d = 10^{-8}$ mg/MeV, $\Lambda_C = 0.5$ MeV-cm$^2$/mg.

The cross-section depends on the tilt angle [32] and reaches the maximum approximately at $\varepsilon_p = 40$ MeV.

The converting procedure (9) from $\sigma(\Lambda)$ to $\sigma_p(\varepsilon_p)$ has been validated by a direct comparison with the heavy ion experimental data of AT68166 [33] and HM628512 [34] memory circuits. Their linear parameters $K_d$ and $\Lambda_C$ has been adopted from [13]. The comparison results are shown in Fig. 5.

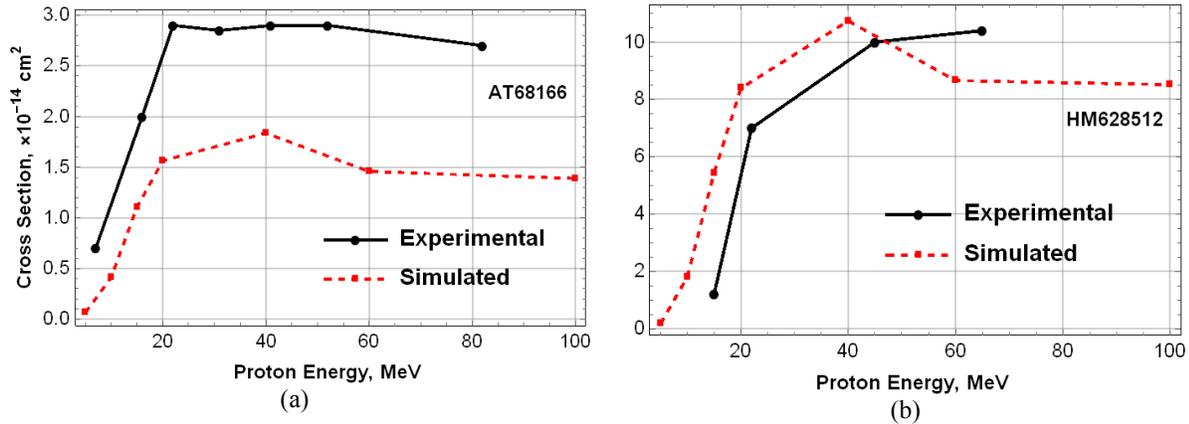

Fig. 5  The simulated (dashed) and experimental (solid) curves $\sigma_P(\varepsilon_p)$ of AT68166 (a) and HM628512 (b) memory ICs.

The simulated curves have a good agreement with data with a maximum discrepancy factor of two.

### 3.3  *Comparison with the in-flight data*

We have been validated our computational approach by a direct comparison with the in-flight data for the Proba-2 [33] and the SAC-C [2] missions. We calculated the LET spectra of secondary particles based on a given proton spectra of the actual orbits. Fig. 6 shows a comparison of the LET spectra for the primary and secondary ions.

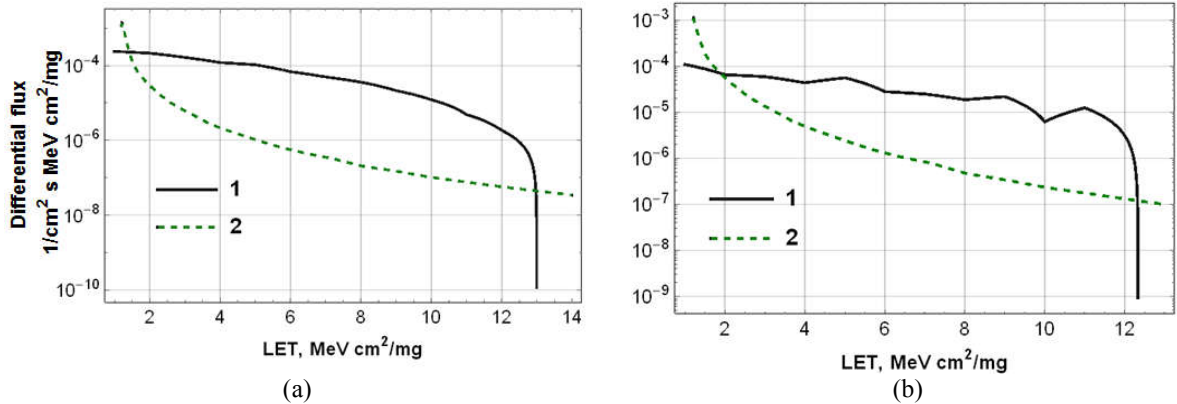

Fig. 6  A comparison of the simulated secondary LET spectra (1) and the primary CREME96 LET spectra (2)
for the Proba-2 (a) and the SAC-C (b) mission orbits.

The simulated spectra $\phi_{SEC}(\Lambda)$ and CREME96 $\phi_{PRIM}(\Lambda)$ were substituted in (1) to calculate the contributions of the primary heavy ions and the proton-induced secondaries to the on-orbit SER.

Table I shows a comparison of in-flight and the calculated SERs for heavy ions and protons of the SRAMs with the known heavy ion cross-section data [2, 35].

TABLE I. SER COMPARISON COMPENDIUM

| Memory circuit | Heavy ions SER (upsets/day) | | Proton SER (upsets/day) | |
|---|---|---|---|---|
| | In-flight | Compact Model Prediction | In-flight | Compact Model Prediction |
| AT68166 | 0.168 | 0.113 | 1.75 | 5.20 |
| HM628512 | 0.05 | 0.07 | 1.03 | 2.49 |
| KM6840003 | 0.35 | 0.73 | 3.59 | 7.43 |
| IS62W20488 | 0.386 | 0.1 | 2.79 | 3.35 |

Simulation results show a good correlation with the in-flight data.

### 3.4 Inverse problem: heavy ion cross sections from proton tests

The proton-induced SEU cross section $\sigma_p^{EXP}$ measured under the test at a given proton energy $\varepsilon_p$ can be, on the other hand, calculated as a result of ionization by the secondary particles (9)

$$\sigma_p^{EXP}(\varepsilon_p) = \int d\Lambda \, \sigma(\Lambda) \, p(\Lambda \mid \varepsilon_p). \tag{14}$$

This relation could be considered as an integral equation for the inverse problem of recovery of $\sigma(\Lambda)$ which is assumed to be unaffected by the primary proton energy. This task is hardly solvable in the general case of an arbitrary form of the $\sigma(\Lambda)$ dependence. However, the problem is essentially simplified for a specific class of the commercial circuits, where the ion cross section can be approximated as a linear function (10). Then (14) takes a following form

$$\sigma_p^{EXP}(\varepsilon_p) = K_d \alpha(\varepsilon_p) \left[ \bar{\Lambda}(\varepsilon_p) - \Lambda_C \right], \tag{15}$$

where $\bar{\Lambda}(\varepsilon_p)$ is defined in (13). The COTS circuits have usually very low critical LETs, typically $\Lambda_C \leq 1$ MeV-cm$^2$/mg. Such low values of $\Lambda_C$ are of order of experimental accuracy of their determination, allowing to fix $\Lambda_C$ at a given low value (say, $\Lambda_C = 0.5$ MeV-cm$^2$/mg). Then, the parameter $K_d$ can be straightforwardly retrieved from (15). We have been applied this simplified procedure to evaluate the heavy ion cross sections based on proton tests.

### 3.5 Validation of the inverse problem

The proposed simplified procedure was verified for the ICs that presumably have the Al overlayers and are consistent with the virtual target in Fig. 1.

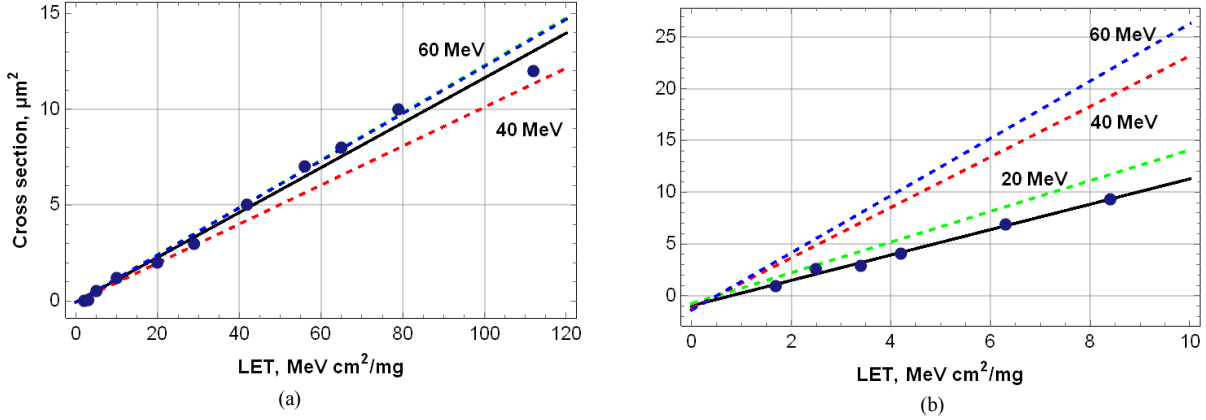

Fig. 7 Direct heavy ion data $\sigma(\Lambda)$ (circles and solid line) and recovered from the proton data (dashed lines) at $\varepsilon_p = $ 20, 40, 60 MeV for the AT68166 (a) and KM6840003 (b) SRAMs. Recovery was performed for a fixed parameter $\Lambda_C$ =0.5 MeV-cm²/mg.

The heavy ion and proton cross section parameters were taken from [9, 33]. Figures 6 show comparison of the experimental and simulated $\sigma(\Lambda)$ for the AT68166 and KM6840003 SRAMs.

The extracted parameters $K_d$ are presented in Table II.

TABLE II. COMPARISON OF SIMULATED AND EXPERIMENTAL HEAVY ION CROSS SECTIONs

| Proton energy $\varepsilon_p$, MeV | IC AT68166 Simulated $K_d$, $10^{-9}$ mg/MeV | | | IC KM6840003, Simulated $K_d$, $10^{-8}$ mg/MeV | | |
|---|---|---|---|---|---|---|
| | Overlayer thickness | | | Overlayer thickness | | |
| | 2.5 um | 22 um | 102 um | 2.5 um | 22 um | 102 um |
| 20 | **1.4** | **1.2** | **1.2** | **1.7** | **1.5** | **1.5** |
| 40 | **1.2** | **1.0** | **1.0** | **2.8** | **2.5** | **2.3** |
| 60 | **1.5** | **1.3** | **1.3** | **3.7** | **2.9** | **3.0** |
| | Experiment $K_d$ = **1.16 $10^{-9}$ mg/MeV** | | | Experiment $K_d$ = **1.22 $10^{-8}$ mg/MeV** | | |

As can be seen from the Fig. 7 and Table II, even the rough virtual target yields a satisfactory estimation of the heavy ion cross section dependence $\sigma(\Lambda)$ for the commercial memory devices.

### 3.6 Fast analytical estimation

Monte-Carlo simulations are quite cumbersome and inconvenient to use. in everyday practice. Simple analytic approximations are very useful in everyday practice to estimate the device SEE susceptibility having only one type of the cross sections. Based on Eq. 15 we propose the following approximate estimation

$$K_d \sim \frac{\sigma_p^{SAT}}{\overline{\alpha} \times (\overline{\Lambda} - \Lambda_C)}, \qquad (16)$$

where $\sigma_p^{SAT}$ is an approximate value of proton-induced SEU cross section, $\overline{\Lambda}$ and $\overline{\alpha}$ are the rigidity and efficacy of the secondary spectra averaged over a range of proton energies. Then, taking the typical values $\Lambda_C \cong 0.5$ MeV-cm²/mg, $\sigma_p^{SAT} \sim 10^{-13}$ cm², $\overline{\Lambda} = 3$ MeV-cm²/mg (see Fig. 3) and $\overline{\alpha} = 5\times10^{-6}$ (see Fig.2), we have a typical value $K_d \cong 10^{-8}$ mg/MeV. We have compared the coefficients $K_d$ recovered from the proton SEU data and independently measured under the ion tests for 4 types of the memory ICs. The results of comparison for the 22 μm overlayer are shown in Table III.

TABLE III. COMPARISON OF CALCULATED AND EXPERIMENTAL HEAVY ION CROSS SECTIONs USING

|  | $K_d$ from the heavy ion data, $10^{-10}$ mg/MeV | $K_d$ recovered from the proton data, $10^{-10}$ mg/MeV |
|---|---|---|
| SRAM 65 nm [36] | 2.9 | 6.4 |
| SRAM 90 nm [37] | 6.2 | 14.4 |
| AT68166 250 nm [38] | 11.6 | 28.0 |
| Commercial RHBD SRAM 90 nm [39] | 21.0 | 32.0 |

This table demonstrates a good correlation between the results and the applicability of a simple analytical approach for rapid recalculation between the two types of experimental data without the use of cumbersome Monte Carlo simulations. A similar numerical correlation between the two types of SEUs was empirically revealed by Petersen in Ref. [40].

### 4. DISCUSSION

Conceptually, there are two distinctly different approaches to the SER prediction. The first, conditionally speaking, dosimetric approach, is based on the concepts of the cell critical energy (or charge), the sensitive volume and the chord lengths distribution in it. Strictly speaking, if we know these parameters (for example, from the circuit simulations), then the dependence of the cross section on the LET is no longer needed. The second approach relies on the phenomenological cross section notion. Actually, if we know the cross section as a function of LET, we do not need to refer to the concepts of the dosimetric approach, since it suffices to use Eq. 1. Indeed, even if we could accurately simulate enormously complicated processes of the charge generation, transport and collection taking into account the specific layout of the IC, we will not be able to get anything more than the cross-section that we have already determined experimentally. The phenomenological SEU cross section does not contain poorly defined microdosimetric parametrs (the sensitive volume sizes, threshold energy) and it can be determined straightforwardly from experiment. The LET spectra of primary and secondary particles are the other necessary constituent for the SER prediction. Monte Carlo approach is an indispensable element for simulation of the nuclear interaction induced LET spectra of secondary particles. We have developed the GEANT4-based simulation tool PRIVET-2 that capable to predict the proton-induced soft error rates based on heavy ion induced testing data, and vice versa. Fig. 8 consolidates the approach we presented in this paper and shows the general flow to evaluate the SER in commercial circuits.

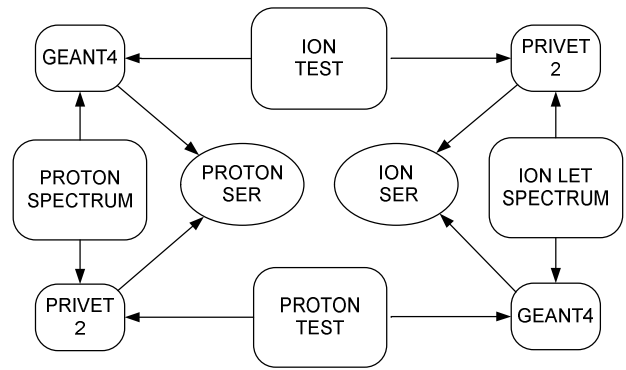

Fig. 8. A flow-chart of the SER simulation procedure using the compact modeling approach.

**Summary**

We have been proposed a simplified procedure to proton and heavy ions induced SER calculation relying only upon the one type of experimental cross-section data (proton or heavy ions) and the spectra on the given orbit. The physical basis for this approach is that in both cases the cause of SEU is direct ionization from the primary or the secondary heavy ions. The proposed GEANT4 assisted approach was validated by direct comparison with the on-board and the accelerator testing results.